\pdfoutput=1

\documentclass[aps,pre,preprint,superscriptaddress]{revtex4-1}

\usepackage[caption=false]{subfig}
\usepackage{graphicx}
\usepackage{amsmath}
\usepackage[hidelinks]{hyperref}
\usepackage[normalem]{ulem}
\usepackage{siunitx}

\usepackage{color}

\begin{document}

\title{Controlled gel expansion through colloid oscillation}

\author{Guido L. A. Kusters}
\email[]{g.l.a.kusters@tue.nl}
\affiliation{Department of Applied Physics, Eindhoven University of Technology, The Netherlands}

\author{Cornelis Storm}
\affiliation{Department of Applied Physics, Eindhoven University of Technology, The Netherlands}
\affiliation{Institute for Complex Molecular Systems, Eindhoven University of Technology, The Netherlands}

\author{Paul van der Schoot}
\affiliation{Department of Applied Physics, Eindhoven University of Technology, The Netherlands}

\date{\today}

\begin{abstract}
We model the behaviour of a single colloid embedded in a cross-linked polymer gel, immersed in a viscous background fluid. External fields actuate the particle into a periodic motion, which deforms the embedding matrix and creates a local micro-cavity, containing the particle and any free volume created by its motion. This cavity exists only as long as the particle is actuated and, when present, reduces the local density of the material, leading to swelling. We show that the model exhibits rich resonance features, but is overall characterised by clear scaling laws at low and high driving frequencies, and a pronounced resonance at intermediate frequencies. Our model predictions suggest that both the magnitude and position of the resonance can be varied by varying the material's elastic modulus or cross-linking density, whereas the local viscosity primarily has a dampening effect. Our work implies appreciable free-volume generation is possible by dispersing a collection of colloids in the medium, even at the level of a simple superposition approximation.

\end{abstract}

\pacs{aaa}

\maketitle

\section{Introduction}
In nature there are many examples of biological systems that must adapt to their environment to maintain functionality, ranging from the intra- and intercellular signalling of proteins \cite{poon2006soft}, to the directed motion, e.g., chemotaxis, of bacteria \cite{wadhams2004making}, and the tailored seed dispersal strategies of plants \cite{elbaum2007role,reyssat2009hygromorphs}. With the envisioned goal of replicating such functionality in mind, scientific research into artificial, stimuli-responsive materials has enjoyed significant interest in recent years \cite{stuart2010emerging,roy2010future,wei2017stimuli,bratek2021towards}. Generally, this concerns systems comprising synthetic polymers that respond suitably to, for example, a variation of the pH or temperature \cite{chen1995graft,schmaljohann2006thermo}, the application of an electric or magnetic field \cite{murdan2003electro,manouras2017field}, or the irradiation of light with a suitable wavelength or intensity \cite{ercole2010photo,gohy2013photo}. Prospective applications include on-demand drug delivery \cite{qiu2001environment,gupta2002hydrogels}, tissue (re)generation \cite{stile2001thermo,miao20174d}, and artificial muscles \cite{shankar2007dielectric,morales2014electro,mirvakili2018artificial}.

The physical properties that must be suitably altered for such applications include, but are not limited to, the porosity \cite{montero2017bioinspired}, conformation \cite{bellomo2004stimuli}, wettability \cite{xin2010reversibly}, and adhesion \cite{boesel2010gecko} of the material. This response is typically programmed into the material on the molecular scale, during the synthesis \cite{stuart2010emerging,roy2010future,wei2017stimuli,bratek2021towards}. Although this enables fine control over the resulting behaviour, it also limits the pool of suitable material candidates. In this paper, we explore an alternative approach, which can render \textit{any} cross-linked polymer network, immersed in a viscoelastic background fluid, responsive in a controlled manner.

To this end, we propose embedding responsive colloids into the gel. Upon subsequent stimulation with an oscillatory external field, e.g., electric or magnetic, we expect the colloids to create microscopic cavities by displacing the embedding matrix during their oscillation. The generated free volume (locally) reduces the density of the material and, in doing so, increases its porosity. 

Although the concept of embedding responsive colloids into a polymer network is not necessarily new \cite{weeber2012deformation,kondaveeti2018magnetic,ganguly2021design}, its application to volume expansion and porosity is. Nevertheless, there is some precedent for the underlying mechanism. For example, Sozanski and co-workers recently conducted viscosity experiments on aqueous solutions of polyethelene glycol using a quartz tuning fork, oscillating at nanoscopic amplitudes \cite{sozanski2014depletion}. For semi-dilute solutions of high molecular weight, they report a significant drop in the local viscosity, orders of magnitude below that of the bulk, which they rationalise based on the motion of a dynamic depletion layer, i.e., a region of free volume. 

In addition, the working mechanism we propose is reminiscent of that utilised in liquid-crystalline networks, where rod-like mesogens are embedded in a dense polymer network \cite{finkelmann1981investigations}. For applications, these mesogens are usually functionalised with either a strong dipole moment \cite{liu2017protruding} or an azobenzene moiety \cite{liu2015new,gelebart2018photoresponsive}. In the former case, application of an alternating electric field induces mesogen reorientation, whereas in the latter case exposure to irradiation of a suitable wavelength induces trans-to-cis interconversion. Both are known to result in the generation of free volume \cite{liu2015new,liu2017protruding}. Similar to the above materials, we envision our approach contributing to the realisation of self-cleaning properties \cite{liu2015reverse}, pattern formation \cite{liu2012photo,mcconney2013topography,babakhanova2019surface,van2019morphing,van2020electroplasticization}, and transport of molecular cargo \cite{cao2019temperature,zhan2020localized}.

To achieve such functionality in a controlled manner, theoretical insight into the underlying mechanism of volume generation is required. To achieve this, we take inspiration from the field of active microrheology \cite{zia2018active}, which likewise studies the driven motion of a colloid immersed in a viscoelastic environment albeit in a non-invasive manner. Since the pioneering work of Batchelor \cite{batchelor1970stress,batchelor1982sedimentation,batchelor1982sedimentation2}, theoretical efforts have primarily focused on relating the interactions between a ``probe" colloid and the local microstructure to global material properties \cite{squires2005simple,khair2006single,zia2010single,hoh2015hydrodynamic,chu2016active}. In addition, the motion of the ``probe" particle itself, as it is driven through the polymer environment, has also been studied extensively, where particular focus has been given to the role of hydrodynamics \cite{odijk2004convective,he2020flow}, depletion interactions \cite{odijk1996protein,odijk2000depletion,tuinier2006depletion,fan2007motion}, and viscosity \cite{kalwarczyk2015motion}. Notwithstanding these important contributions, our focus is instead on how the motion of such a colloid can \textit{perturb} the local microstructure. We use a considerably simpler model in the form of a driven harmonic oscillator, extended to reflect its coupling to the viscoelastic environment, i.e., the cross-linked polymer gel with viscous background fluid, into which the colloid is embedded.

We detail our theoretical model, based on a generalised harmonic oscillator, in Secs. \ref{sec:model}-\ref{sec:scaled model}.
Following this, in Sec. \ref{sec:resonance}, we characterise the oscillation of the colloid by reporting the amplitude as a function of the driving frequency. We show numerically that the model exhibits clear scaling laws in the low- and high-frequency limits, and a resonance at intermediate driving frequencies. These can all be calculated reasonably accurately by making simplifying assumptions. 
Next, in Sec. \ref{sec:volume}, we link the colloid oscillation to volume expansion by investigating the steady-state cavity volume. For this we likewise derive clear scaling laws and locate a distinct resonance that suggests volume expansion far exceeding the length scale of the colloid is possible.
Subsequently, in Sec. \ref{sec:params}, we investigate the effect of varying both the elasticity of the polymer network, accessible via the gel's elastic modulus or cross-linking density, and the viscosity of the background fluid. The former shifts both the position and the magnitude of the resonance, whereas the latter mainly has a dampening effect. Finally, we conclude by summarising our most salient results and reflecting on their significance in view of experiments, in Sec. \ref{sec:conclusion}.

\section{Theoretical model}\label{sec:model}
We model the motion of a colloid embedded in a cross-linked polymer gel, which is in turn immersed in a viscous background fluid, by means of a generalised, driven harmonic oscillator. As point of departure, we consider a colloid of mass $m$ and diameter $2a$, which we assume is greater than the distance between permanent crosslinks in the gel. The colloid is driven through a viscoelastic environment that is characterised by an elastic spring constant $k$ and a viscous damping coefficient $c$. For a gel the latter is directly related to the viscosity of the background fluid. Assuming a driving force of magnitude $F_0$ and driving frequency $\omega$, the resulting equation of motion for the colloid position $x=x\left(t\right)$ reads
\begin{equation}\label{eq:DHM}
    m\ddot{x}+c\dot{x}+kx=F_0\sin\omega t,
\end{equation}
where dots indicate derivatives with respect to time $t$. For the remainder of this paper we assume that, initially, the colloid is at rest and there is no residual stress in the network.

\begin{figure*}[htbp]
    \subfloat{\includegraphics[width=\textwidth]{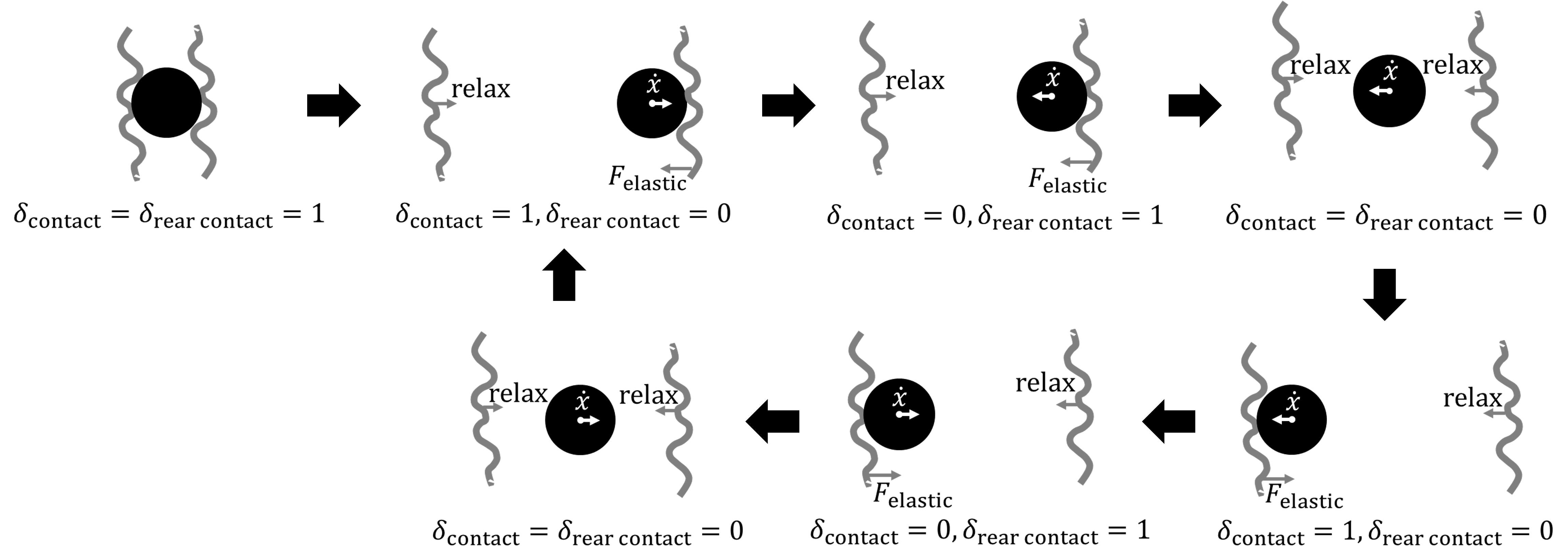}}\\
    \caption{Schematic illustration of the colloid oscillation, including the direction of the velocity, forces, and polymer relaxation. The significance of the functions $\delta_\text{contact}$ and $\delta_\text{rear contact}$, discussed in the main text, is also illustrated.}\label{fig:cartoon}
\end{figure*}

The philosophy of the model is that the colloid, through its oscillatory motion, creates a microscopic cavity, containing itself and the viscous background fluid, by displacing polymeric material. Fig. \ref{fig:cartoon} schematically shows this process. 
From the first two panels we see that, as the external field is applied, the colloid perturbs the polymer network from its initial, stress-free configuration by locally compressing it. This compression is resisted by an elastic restoring force, and the cavity that opens up in the wake of the colloid relaxes viscoelastically. The third panel coincides approximately with the change in sign of the oscillatory external field, marking the turning point of the oscillation. Here, the elastic restoring force, which is now directed parallel to the colloid velocity, still applies and the viscoelastic relaxation of the cavity likewise continues. 
Following this, the fourth panel illustrates that if the colloid separates from the cavity wall, no elastic force acts on it, and both cavity walls relax viscoelastically. Finally, the bottom panels show how the oscillation continues as the colloid makes contact with the opposite cavity wall, and the same steps are repeated. The colloid experiences friction with the viscous background fluid at all of these stages.

For the sake of simplicity, we have tacitly assumed that the polymer network does not screen the viscous friction of the colloid with the background fluid in writing the above, such that only the elastic force depends on whether the colloid makes contact with the cavity wall. This transforms Eq. \eqref{eq:DHM} into
\begin{equation}\label{eq:DHM2}
    m\ddot{x}+c\dot{x}+kx \, \delta_\text{contact}=F_0\sin\omega t.
\end{equation}
where we have introduced the ``contact function," $\delta_\text{contact}$, which is unity if the colloid is in contact with a cavity wall
and zero otherwise (see Fig. \ref{fig:cartoon}).
Note that our qualitative results in this paper are robust even upon relaxing the assumption of no screening, which we explore in the Supplementary Material.

The above makes clear that physical contact between the colloid and cavity wall is a crucial aspect of the model. However, to realise its consistent implementation, we must make a choice regarding the equilibrium configuration the polymer network relaxes toward. Is all relaxation in our one-dimensional description purely elastic, with any given cavity wall only relaxing as far as its initial position, or can plastic rearrangements also occur, where the cavity wall relaxes beyond its initial position to fill the microscopic cavity? In this paper, we pursue the latter choice, as we intuitively expect a unilateral displacement of the colloid to not give rise to any \textit{lasting} free volume. In addition, if we were to for the moment consider the free-volume generation from a three-dimensional perspective, not only the cavity walls but also the sides of the cavity play a role. That is, even if the non-compressed cavity wall is under no stress, the sides of the cavity experience extensional stresses that promote locally shrinking the cross section of the cavity. Translated to our one-dimensional picture, this again results in a relaxation of the cavity wall that effectively goes beyond its initial position. Thus, in what follows, we assume that the microscopic cavity relaxes viscoelastically until it fully collapses. Further below we explain how we model this.

However, we first address a complication this choice carries with it. Namely, if the cavity wall is allowed to relax beyond its initial position via plastic rearrangements, we must give careful thought to the rest position of the elastic force, $x_{0,\text{wall}}\left(t\right)$, and how it evolves with time; this rest position is generally different for the different cavity walls, meaning $x_{0,\text{left}}\left(t\right)\neq x_{0,\text{right}}\left(t\right)$. We cannot simply set $x_{0,\text{wall}}\left(t\right)=0$, since if the cavity wall in question relaxes beyond its initial position $x=0$, it would then not resist but instead accelerate the motion of the colloid upon contact. Equating the rest position, $x_{0,\text{wall}}\left(t\right)$, with the colloid position, $x\left(t\right)$, at any time that contact is established leads to different complications. This approach would artificially and progressively weaken the network in the case the external field repeatedly pushes the colloid into the polymer network, which in turn rebuffs it; this is a scenario we encounter in our numerical evaluation of the model.

We reconcile the above concerns by, upon establishing contact between colloid and cavity wall, \textit{only} equating the rest position with the colloid position if the colloid has moved beyond the current rest position, i.e., if $x\left(t\right)>x_{0,\text{left}}\left(t\right)$ or $x\left(t\right)<x_{0,\text{right}}\left(t\right)$. In essence, this means we assume relaxation is effectively elastic until the cavity wall reaches its current rest position. Beyond this point, relaxation occurs via plastic rearrangements and thus continually increases the rest position, until the cavity wall is again perturbed by the colloid. We carry out this procedure separately for both cavity walls. With this in mind, Eq. \eqref{eq:DHM2} becomes
\begin{equation}\label{eq:DHM3}
    m\ddot{x}+c\dot{x}+k\left(x-x_{0,\text{wall}}\right)\delta_\text{contact}=F_0\sin\omega t,
\end{equation}
where $x_{0,\text{wall}}=x_{0,\text{wall}}\left(t\right)$ depends crucially on the cavity wall with which the colloid is in contact and the deformation history of the network. A consequence of this choice is that inversion symmetry along the $x$-axis is broken: applying an external field $\propto\pm\sin\omega t$ yields identical oscillations, but mirrored in $x=0$. Thus, within this set of assumptions, the transient stages of dynamics can have a lasting effect on the eventual steady state that is reached: a direct consequence of the occurrence plastic rearrangements. This we shall see when we discuss our results in Sec. \ref{sec:resonance}.

Finally, we link the colloid dynamics to that of the microscopic cavity by specifying the evolution of the cavity volume. To this end, we propose that the colloid creates a tubular cavity, with a cross section equal to that of the colloid, during its oscillation. The corresponding increase in cavity free volume $v$, i.e., the volume of the cavity in excess of the colloid volume, then reads
\begin{equation}\label{eq:vol}
    \dot{v}=\lvert\dot{x}\rvert \pi a^2\, \delta_\text{contact}-\frac{v}{\tau},
\end{equation}
where the first term, representing expansion, only applies if the colloid is in contact with the polymer network. The absolute sign implies that variation in cavity free volume occurs at double the frequency of the colloid oscillation, as free volume is generated upon both leftward and rightward motion. The second term describes the relaxation of the microscopic cavity, with $\tau$ the typical relaxation time of the cross-linked polymer gel. The underlying assumption here is one of Rouse-like relaxation, where we take the dynamics to be dominated by the lowest-order mode \cite{rouse1953theory}. That is, we assume the slowest mode that can be excited is on the order of the distance between permanent crosslinks, which we associate with the relaxation time $\tau$, and we neglect modes on shorter length scales, which relax significantly faster. Looking ahead, relaxation of deformations on longer length scales, such as for relaxation of the bulk, can still occur on longer time scales. However, as we shall see below, the system eventually achieves a steady state oscillation, such that, if given enough time, the bulk can invariably adjust to any local changes in density effected by the colloid.

This concludes the discussion of our theoretical model. Before we investigate the resulting model dynamics, we first scale the theory to make it dimensionless, reducing the number of variable parameters.

\section{Scaling procedure}\label{sec:scaled model}

To scale equations \eqref{eq:DHM3} and \eqref{eq:vol}, we take $\tau$ as the typical time scale of the host material, and identify $L=F_0\tau^2/m$ as its typical length scale. This quantity is physically related to the distance a force of magnitude $F_0$ would pull the colloid of mass $m$ in the relaxation time $\tau$, if we disregard the embedding matrix and background fluid. Subsequently introducing the scaled time, $T=t/\tau$, the scaled colloid position, $X=x/L$, and the scaled cavity volume, $V=v/L\pi a^2$, the set of scaled governing equations becomes
\begin{equation}\label{eq:scaled xV}
    \begin{cases}
        &\ddot{X}+2B\dot{X}+\Omega_0^2\left(X-X_0\right)\delta_\text{contact}=\sin\Omega T,\\
        &\dot{V}=\lvert{\dot{X}}\rvert\, \delta_\text{contact}-V.
    \end{cases}
\end{equation}
Here, the derivatives are taken with respect to the scaled time, and we have expressed the model in terms of the scaled quantities
\begin{equation}
    \begin{cases}
        B&=\frac{c\tau}{2m},\\
        \Omega_0&=\sqrt{\frac{k}{m}}\tau\\
        \Omega&=\omega\tau.
    \end{cases}
\end{equation}

The above procedure reduces the number of model parameters to two: the scaled damping ratio, $B$, and the scaled natural frequency of the oscillator, $\Omega_0$. The most straightforward experimental link to the former is through the viscosity of the background fluid, whereas the latter depends strongly on the gel's elastic modulus and its cross-link density. Also implicit in the scaling is that our results can be tuned by varying the mass and radius of the colloid, the relaxation time of the gel, and the magnitude of the external driving force. The above allows us to comprehensively scan the parameter space, and numerically find the steady-state solutions for a given scaled driving frequency, $\Omega$. This we do by means of a standard fourth-order Runge-Kutta scheme, adapted to account for the contact function we use.

This establishes the scaled model we use for the remainder of this paper. In order to provide an overview of the dynamical response it gives rise to, we now present the ensuing colloid dynamics for a range of driving frequencies.

\section{Colloid dynamics}\label{sec:resonance}

\begin{figure}[htbp]
    \subfloat[]{\includegraphics[width=0.49\textwidth]{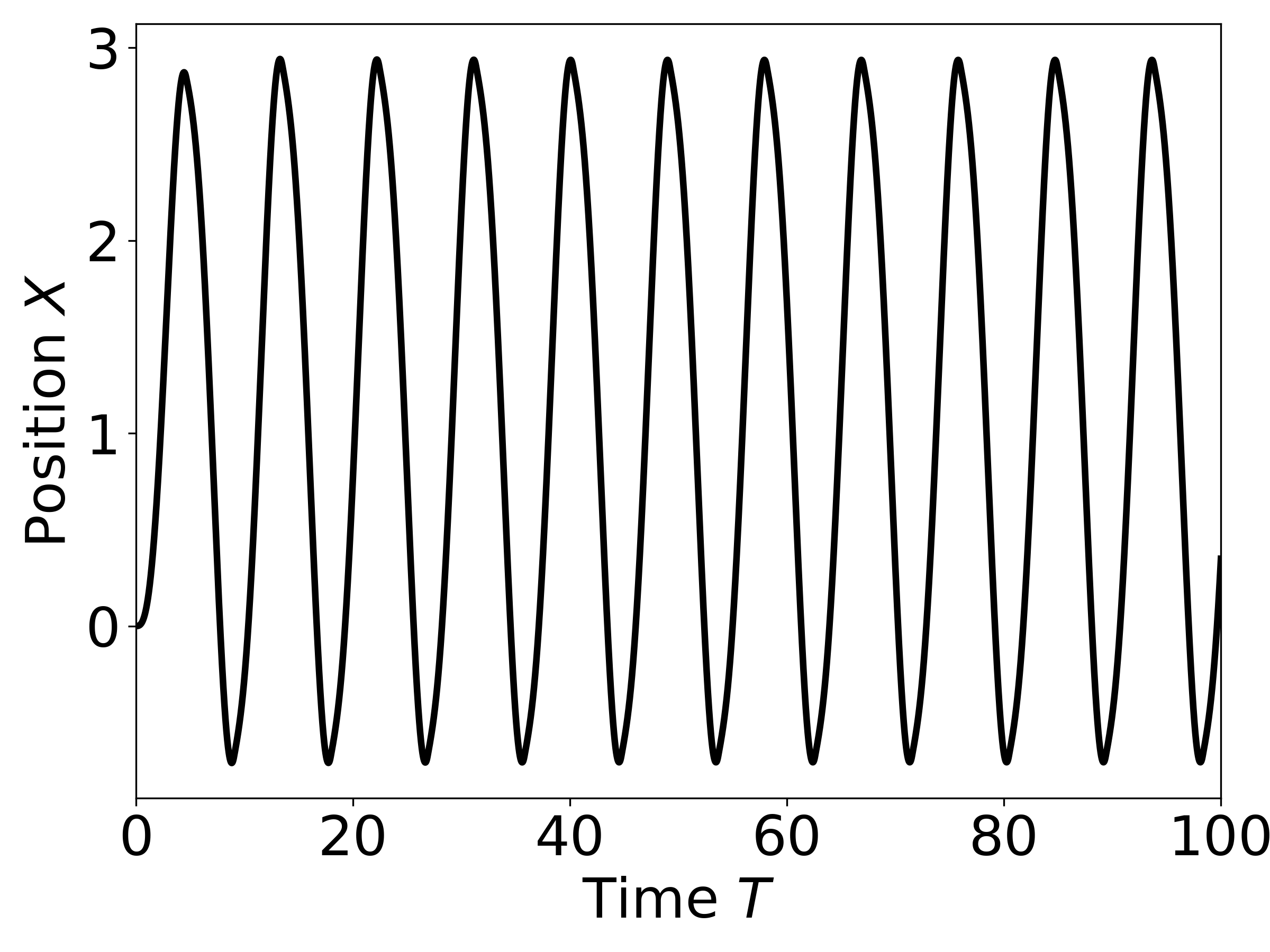}}
    \subfloat[]{\includegraphics[width=0.49\textwidth]{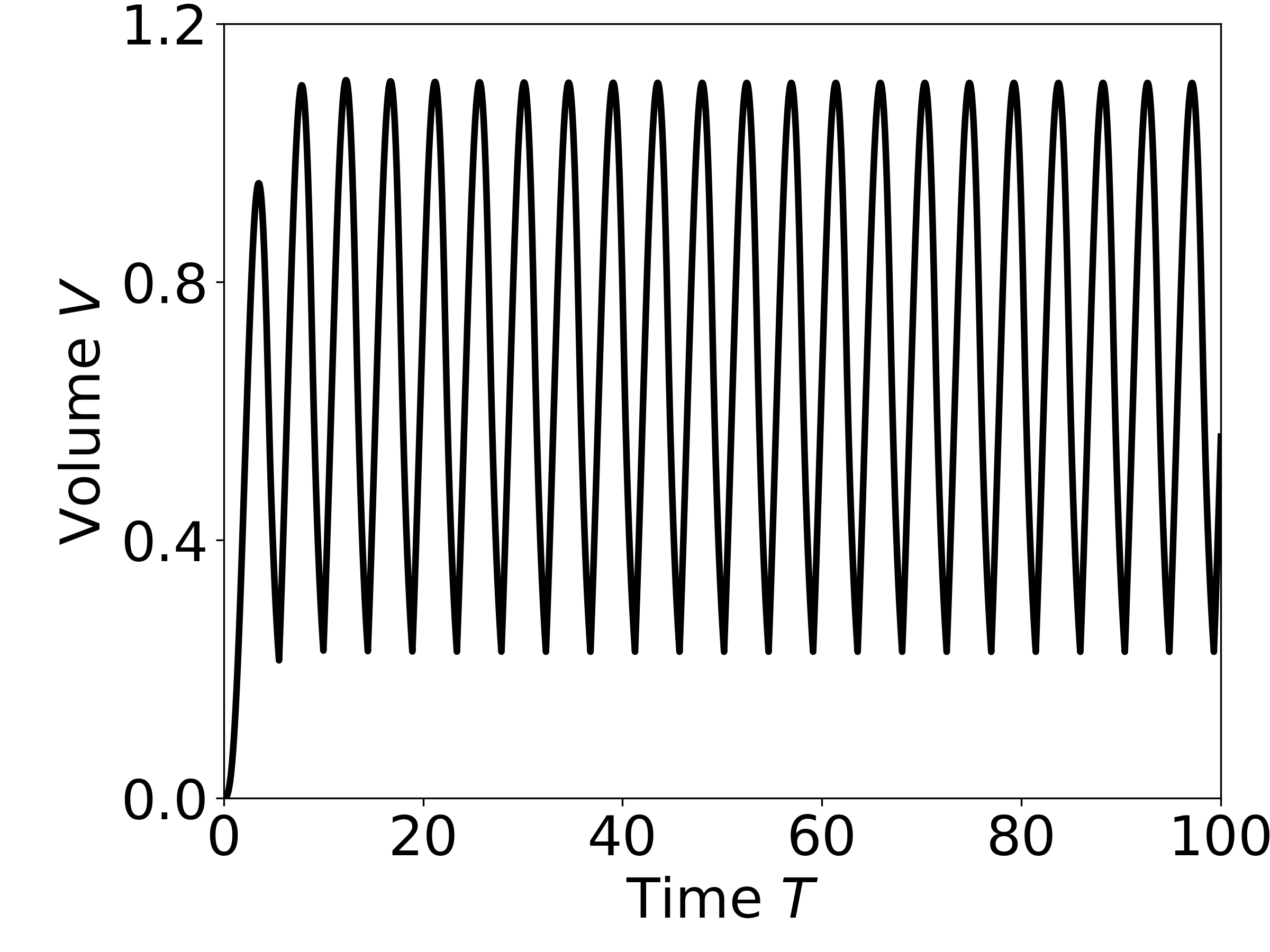}}\\
    \subfloat[]{\includegraphics[width=0.475\textwidth]{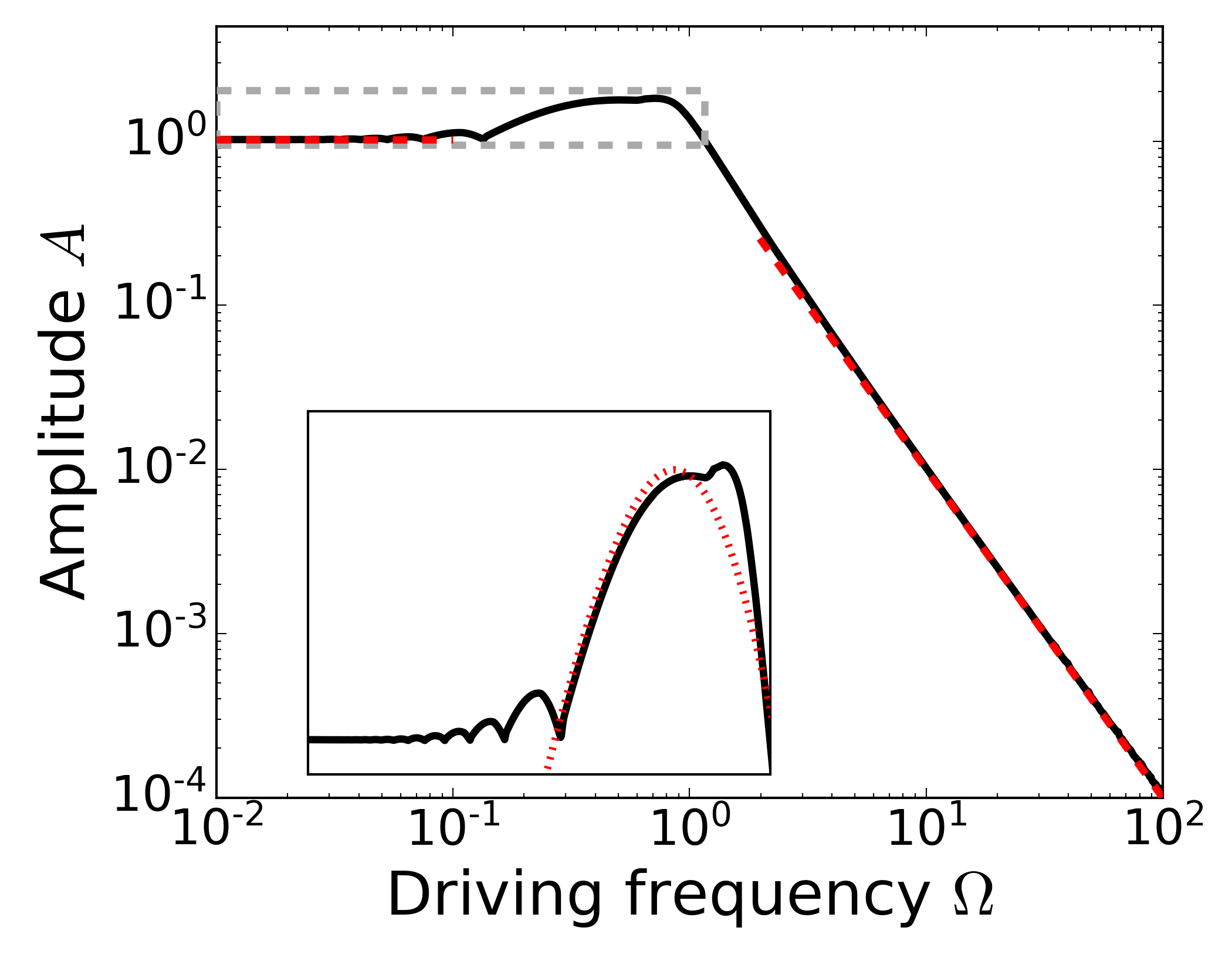}}
    \subfloat[]{\includegraphics[width=0.49\textwidth]{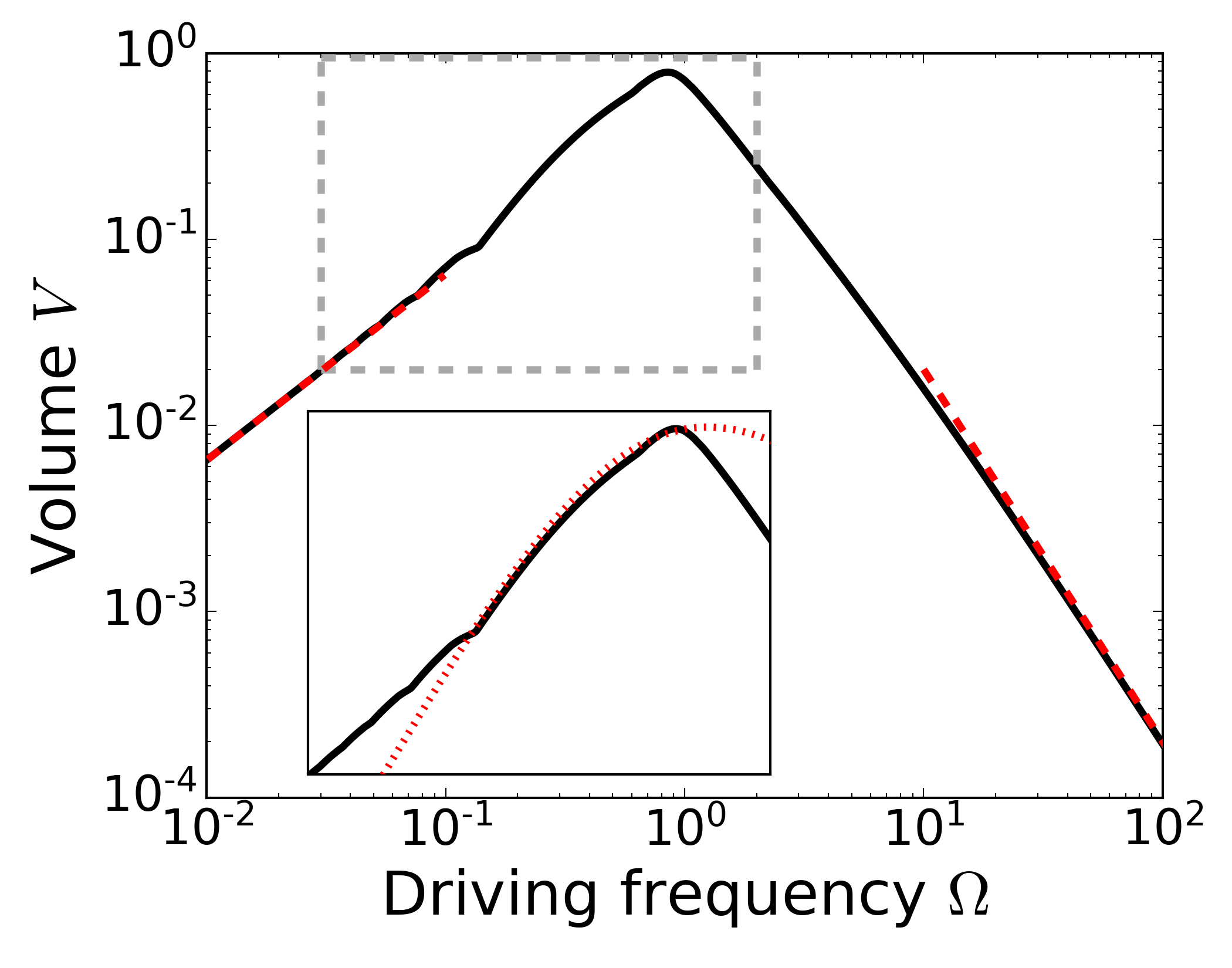}}\\
    \caption{Scaled steady-state amplitude, $A$, of the colloid oscillation (left) and the corresponding cavity volume, $V$, (right) as a function of the scaled driving frequency $\Omega$, on logarithmic scale. The dashed red lines denote the low- and high-frequency scaling laws $A_\text{low}=1/2\Omega_0^2$ and $V_\text{low}=\Omega/\pi\Omega_0^2$, and $A_\text{high}=1/\Omega^2$ and $V_\text{high}=2/\Omega^2$, respectively. The zoomed-in inset is indicated in grey; the dotted red curves in the insets correspond to the theoretical estimates of Eq. \eqref{eq: amp estimate} (left) and Eq. \eqref{eq: vol estimate} (right), respectively. Parameter values used: $\Omega_0=0.7$, $B=0.05$.}\label{fig:resonance}
\end{figure}

Fig. \ref{fig:resonance}(a) shows a typical oscillation of the colloid position, corresponding to a driving frequency $\Omega=\Omega_0/\sqrt{1-2B^2/\Omega_0^2}$. This is the resonant frequency of the ``bare" harmonic oscillator, in the absence of the changes we introduced to model the viscous cavity. From the figure we see that, unlike what is expected for a simple harmonic oscillator, the colloid traces out a trajectory that is asymmetric about the origin. As foreshadowed, this is a direct consequence of allowing plastic rearrangements within the model, as opposed to purely elastic deformations. Since which plastic rearrangements occur is largely determined during the transient stages of dynamics, the initial form of the external field matters for the eventual steady state that is reached. 

This also explains why we deliberately choose to apply a sine function as the external field, rather than a cosine. The initial ``kick" provided by the latter would immediately induce plastic rearrangements, which in turn affect the steady state. This presents a counter-intuitive prediction---the distinctness of the steady-state oscillations driven by sine and cosine fields, especially at low frequencies---that could be straightforwardly verified through experiments. In the SM we work out the corresponding differences and their origins, which we expect to become more important for more densely packed and cross-linked polymer networks. In this paper, however, we shall err on the side of caution by focusing on external fields that initially vary slowly, so as to prevent possibly spurious plastic rearrangements. We return to this point at the end of Sec. \ref{sec:conclusion}.

To investigate the dynamics of the colloid from a broader perspective, Fig. \ref{fig:resonance}(c) shows the scaled steady-state oscillation amplitude of the colloid, $A$, as a function of the driving frequency of the external field, $\Omega$. From this we primarily conclude that the model obeys clear scaling laws in the low- and high-frequency limits, as indicated by the dashed red lines. In between these limits, at intermediate driving frequencies, we find a pronounced resonance. This is clarified by the zoomed-in inset, the boundaries of which are indicated by the dashed grey lines. Here, the dotted red line represents a theoretical estimate, which we shall discuss further below (see Eq. \eqref{eq: amp estimate}). Finally, the model also exhibits a host of finer features, including the resonance peak being split in two and the emergence of multiple local optima in the low-frequency regime. These features add to the depth of our model but are not crucial to our message in this paper, hence we discuss them in the Supplementary Material.


We commence by rationalising the amplitude in the low-frequency limit, where we expect only the elastic and driving force in Eq. \eqref{eq:scaled xV} to contribute meaningfully to the dynamics. As a result, the two should balance at the turning point of the oscillation, where $\sin\Omega T=1$, yielding $A_\text{low}=1/2\Omega_0^2$ (dashed red line). From the figure it is apparent there is good agreement with the numerical results (black curve). Conversely, we expect the high-frequency limit to reflect a competition between the inertial term and the driving force. Equating the two and Fourier transforming the result shows that steady-state solutions must take the form $X\left(t\right)=\sin\Omega T/\Omega^2$, which likewise agrees with the scaling law $A_\text{high}=1/\Omega^2$ shown in the figure (dashed red line).

Although neither of the above lines of reasoning translates perfectly to intermediate driving frequencies, here we can make headway by neglecting viscous friction and presuming that the colloid maintains contact with the cavity wall until reaching its turning point. This is reasonable for systems dominated by elastic rather than viscous effects. If we furthermore presume that the colloid leaves this turning point with no significant residual velocity or acceleration, it follows the same trajectory back to its starting position, from which the oscillation continues identically. Under these assumptions, the entire oscillation is effectively determined upon reaching the first turning point, suggesting that its amplitude can be deduced by computing said turning point. This we do by setting $\delta_\text{contact}=1$ and $B=0=X_0$ in Eq. \eqref{eq:scaled xV}, which yields
\begin{equation}\label{eq: amp estimate}
    A_\text{intermediate}=\frac{1}{2}\frac{\Omega  \sin \left(\frac{2 \pi  \Omega_0}{\Omega
   +\Omega_0}\right)-\Omega_0 \sin \left(\frac{2 \pi
    \Omega }{\Omega +\Omega_0}\right)}{\Omega^2
   \Omega_0-\Omega_0^3}.
\end{equation}
This equation is plotted as a dotted red line in the inset of Fig. \ref{fig:resonance}(c), from which good agreement with the numerical results is apparent for the modest values of $\Omega_0$ and $B$ used here. Below, in Sec. \ref{sec:volume}, we derive a similar equation for the concomitantly generated free volume $V$.

The above estimate and scaling relations capture the overall response of the colloid to the external field. Below, in Sec. \ref{sec:volume}, we discuss the overall cavity volume the colloid oscillation gives rise to in the steady state.

\section{Cavity expansion}\label{sec:volume}

Fig. \ref{fig:resonance}(b) shows the cavity free volume, i.e., the cavity volume in excess of the colloid volume, resulting from the colloid oscillation shown in Fig. \ref{fig:resonance}(a). The frequency of this curve is twice as high as that of the colloid oscillation, since free volume is generated upon both leftward and rightward motion. In addition, the asymmetry between the bottom (sharp) and the top (rounded) of the curve can be traced back to free volume generation being tied directly to the colloid making contact with the cavity wall.

To investigate how we might influence the cavity dynamics through our choice of external actuation, Fig. \ref{fig:resonance}(d) shows the steady-state cavity free volume for a range of driving frequencies. 
The figure is closely related to the steady-state oscillation amplitude shown in Fig. \ref{fig:resonance}(c), which can be interpreted as an upper bound for free volume generation. The actual free volume is lower due to the viscoelastic relaxation of the gel.


As before, the model exhibits distinct scaling laws in the low- and high-frequency limits, indicated by the dashed red lines. Here, the low-frequency behaviour is markedly different from that of the oscillation amplitude (Fig. \ref{fig:resonance}(c)) due to the aforementioned relaxation, tending to zero in the low-frequency limit. The zoomed-in inset shown in the figure again highlights the resonance of the system, where the dotted red curve denotes a theoretical estimate to which we shall return further below (see Eq. \eqref{eq: vol estimate} below). We discuss the finer features of the model in the Supplementary Material.

Notably, Fig. \ref{fig:resonance}(d) communicates that a steady-state expansion of up to $V\approx1$, as measured on the system's natural length scale $L$, can be achieved.
We argue that this constitutes a significant expansion, which we support by means of a rough estimate. If we assume a homogeneous, spherical gold nanoparticle with a radius of $a=\SI{25}{\nano\meter}$ and a surface charge density of $-\SI{2}{\milli\coulomb}$ \cite{kumal2015determination}, subject to an electric field with strength $\SI{1}{\volt/\meter}$ and immersed in a polystyrene network with Rouse time $\tau=\SI{1}{\centi\second}$ \cite{west1969relaxation}, we find that the system's typical length scale equals $L\approx\SI{1}{\milli\meter}$. Given that the resonance peak shown in Fig. \ref{fig:resonance}(d) obeys $V\approx1$, measured on this length scale, we conclude that volume increases far exceeding the length scale of the colloid are in principle possible.

Next, we validate the model through scaling relations in the low- and high-frequency limits. For low driving frequencies, we expect that the colloid is continually pushing against the cavity walls, and that in the steady state the concomitant expansion is compensated exactly by the viscoelastic relaxation of the gel (see Eq. \eqref{eq:scaled xV}). Thus, estimating the average velocity of the colloid as $\left(1/\Omega_0^2\right)/
\left(\pi/\Omega\right)$ by using the low-frequency estimate of the oscillation amplitude we find $V_\text{low}=\Omega/\pi\Omega_0^2$, as indicated by the left-most dashed red line in Fig. \ref{fig:resonance}(d).

Conversely, for high driving frequencies we expect that the cavity walls have no time to relax, suggesting that in the steady state the extent of the cavity equals twice the oscillation amplitude, $V_\text{high}=2/\Omega^2$. This is shown by the right-most dashed red line in Fig. \ref{fig:resonance}(d), which again demonstrates good agreement between the expected scaling relations and the model. 

Finally, for intermediate driving frequencies, we already approximated the oscillation amplitude (Eq. \eqref{eq: amp estimate}) by neglecting viscous friction and presuming constant contact between colloid and network. Consequently, we can estimate the steady-state cavity volume using the same set of approximations by setting $\lvert\dot{X}\rvert\approx2A_\text{intermediate}/\left(\pi/\Omega\right)$ in Eq. \eqref{eq:scaled xV}. This yields
\begin{equation}\label{eq: vol estimate}
    V_\text{intermediate}=\frac{\Omega}{\pi}\frac{\Omega  \sin \left(\frac{2 \pi  \Omega_0}{\Omega
   +\Omega_0}\right)-\Omega_0 \sin \left(\frac{2 \pi
    \Omega }{\Omega +\Omega_0}\right)}{\Omega^2
   \Omega_0-\Omega_0^3},
\end{equation}
which is indicated by the dotted red line in the inset of Fig. \ref{fig:resonance}(d). The agreement with the full model is again good for the used parameter values.

This establishes the general behaviour of the expanding cavity. We now investigate the effect of varying model parameters.

\section{Varying model parameters}\label{sec:params}

\begin{figure}[htbp]
    \subfloat[]{\includegraphics[width=0.475\textwidth]{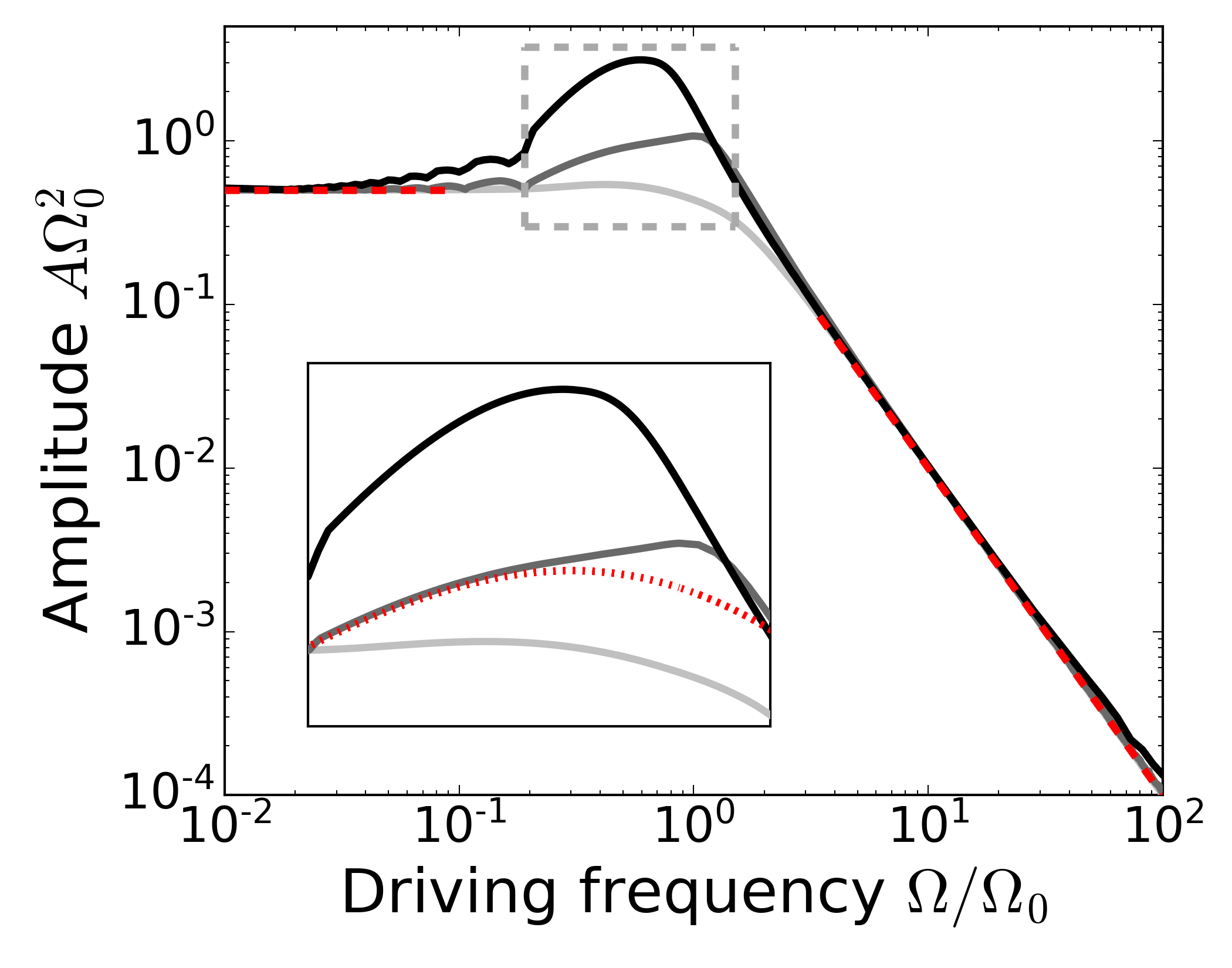}}
    \subfloat[]{\includegraphics[width=0.49\textwidth]{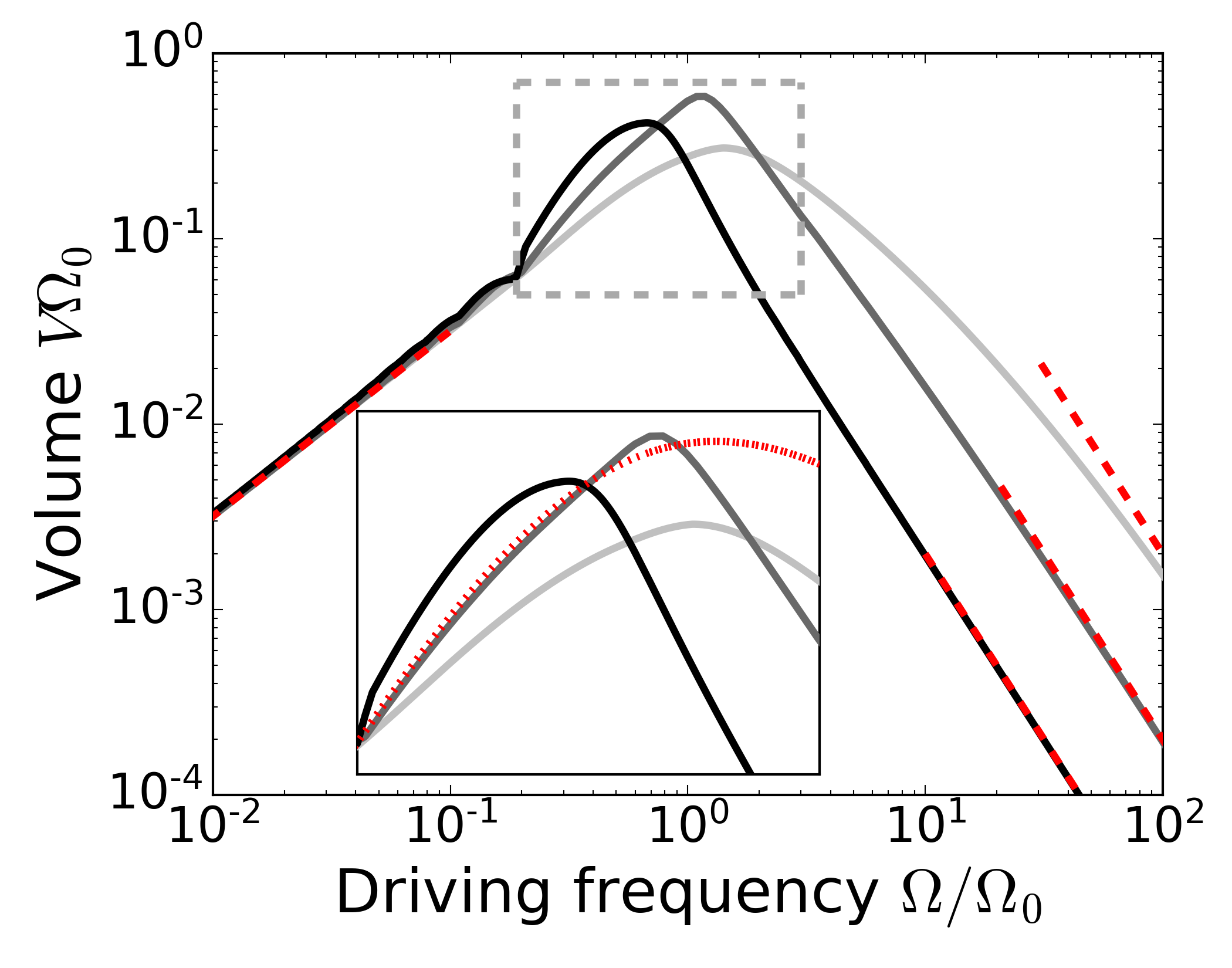}}\\
    \caption{Scaled steady-state amplitude, $A$, of the colloid oscillation (left) and the corresponding cavity free volume, $V$, (right) as a function of the scaled driving frequency $\Omega$, on logarithmic scale. The curves correspond to $\Omega_0=0.1$ (silver), $\Omega_0=1.0$ (grey), and $\Omega_0=10.0$ (black), all for $B=0.05$. The dashed red lines denote low- and high-frequency scaling laws. The zoomed-in inset is indicated in grey; the dotted red curves in the insets correspond to the theoretical estimates of Eq. \eqref{eq: amp estimate} (left) and Eq. \eqref{eq: vol estimate} (right), respectively.}\label{fig:resonancew0}
\end{figure}

The key parameters of interest are the (scaled) natural frequency of the network, $\Omega_0$, closely related to the elastic modulus and cross-linking density, and the (scaled) damping ratio, $B$. Fig. \ref{fig:resonancew0}(a) shows the steady-state oscillation amplitude as a function of the driving frequency, for $\Omega_0=0.1$ (silver curve), $\Omega_0=1.0$ (grey curve), and $\Omega_0=10.0$ (black curve). All curves correspond to the same damping ratio $B=0.05$.
Here, we have taken care to scale the amplitude $A\Omega_0^2$ and the driving frequency $\Omega/\Omega_0$, such that the different lines collapse onto a single curve in the low- and high-frequency limits. This already implies that the position of the resonance can be shifted by varying $\Omega_0$ and that, similarly, its amplitude scales inversely with $\Omega_0$. That is, a greater elastic modulus both speeds up the oscillation, and suppresses its amplitude. Any additional features apparent from Fig. \ref{fig:resonancew0}(a) are superimposed on top of this dominating trend.

In particular, the inset of Fig. \ref{fig:resonancew0}(a) shows that the behaviour of the resonance peak itself is more subtle than our introduced scaling implies: here the different curves don't overlap. As it turns out, the magnitude of this peak depends more strongly on $\Omega_0$ than the $\propto\Omega_0^{-2}$ dependence predicted by the theoretical estimate of Eq. \eqref{eq: amp estimate}, which is shown in the inset by the dotted red curve. This dependence is already absorbed into the chosen scaling. Although the qualitative agreement with all curves is reasonable, this estimate only \textit{quantitatively} describes the grey curve, which corresponds to the (scaled) natural frequency $\Omega_0=\tau\sqrt{k/m}=1.0$. In other words, our theoretical estimate for the resonance is quantitative if the natural time scale of the oscillation, $\sqrt{k/m}$, is comparable to the relaxation time of the polymer gel, $\tau$, and becomes qualitative if one grows much larger than the other.


The above suggests that the assumptions we make in deriving Eq. \eqref{eq: amp estimate} break down for exceedingly weak or strong elastic moduli. For the former, the elastic restoring force no longer dominates the viscous friction with the background fluid, which in turn slows down the colloid and diminishes the amplitude of the oscillation more than predicted theoretically. For the latter, the elastic restoring force is sufficiently strong such that the colloid does not leave the initial turning points of its oscillation with negligible residual velocity. Instead, the oscillation grows in amplitude over the first few periods before settling to a steady state.
As a result, the theoretical estimate underestimates the steady-state amplitude in this case.

\begin{figure}[htbp]
    \subfloat[]{\includegraphics[width=0.475\textwidth]{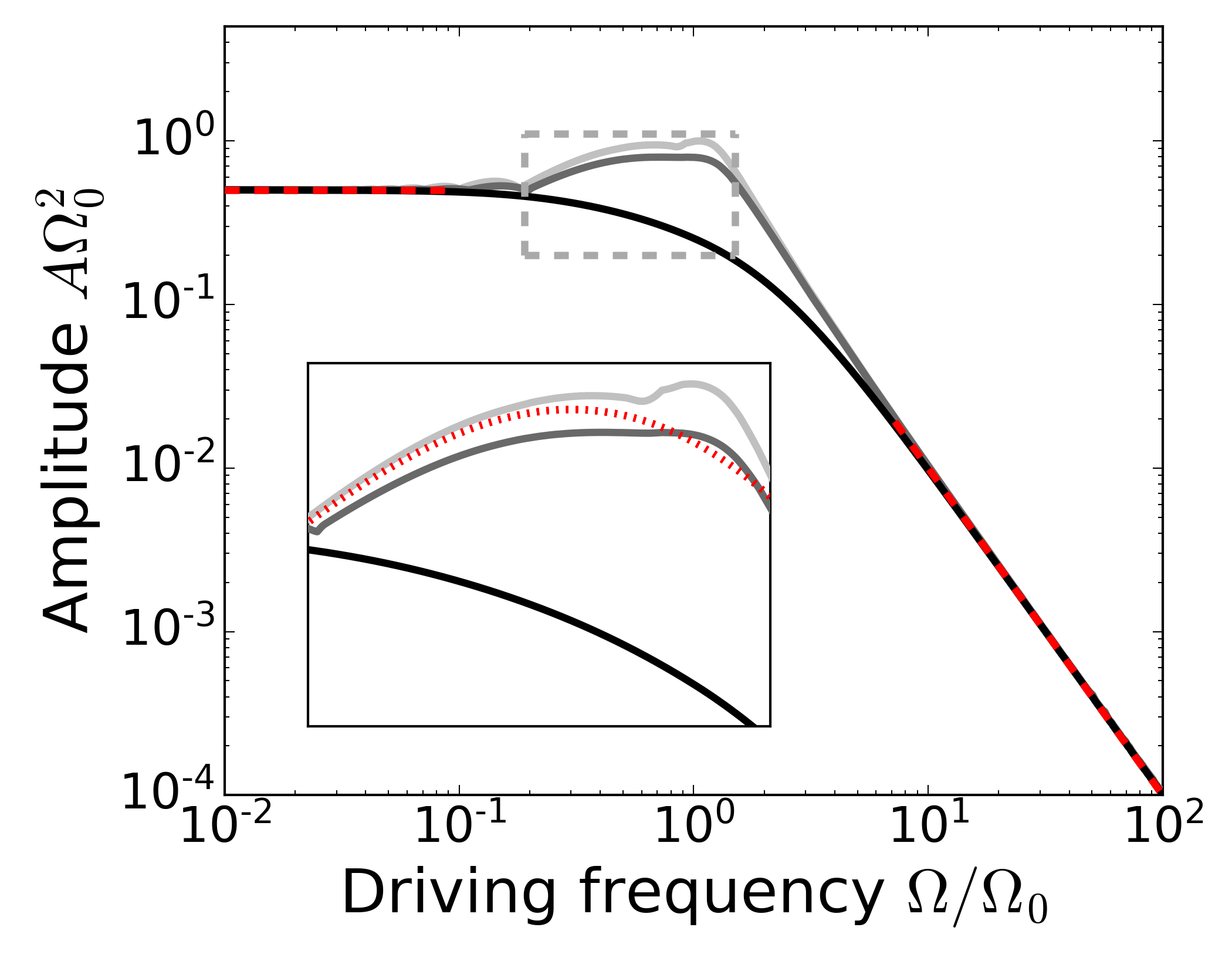}}
    \subfloat[]{\includegraphics[width=0.49\textwidth]{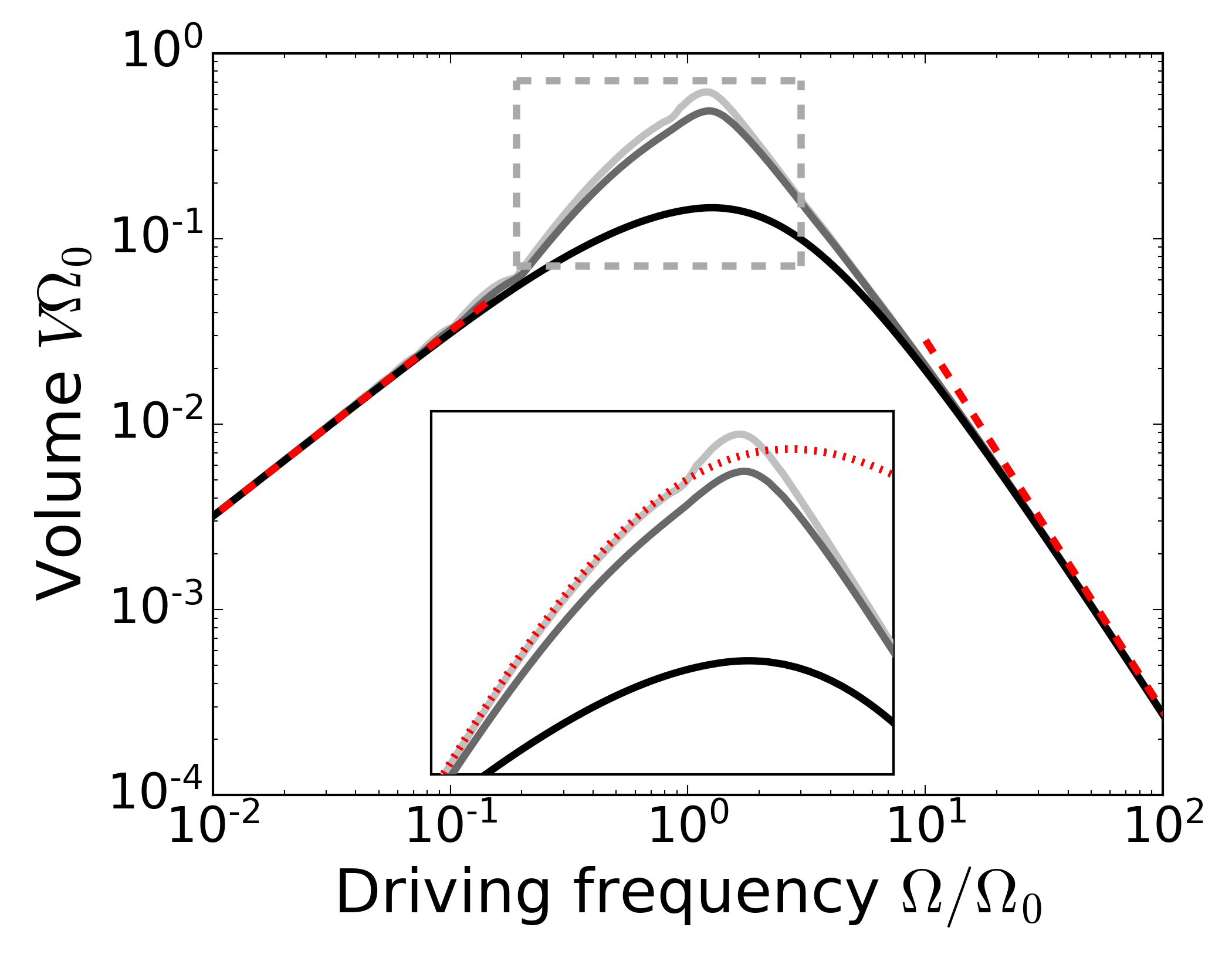}}\\
    \caption{Scaled steady-state amplitude, $A$, of the colloid oscillation (left) and the corresponding cavity free volume, $V$, (right) as a function of the scaled driving frequency $\Omega$, on logarithmic scale. The curves correspond to $B=0.01$ (silver), $B=0.1$ (grey), and $B=1.0$ (black), all for $\Omega_0=0.7$. The dashed red lines denote low- and high-frequency scaling laws. The zoomed-in inset is indicated in grey; the dotted red curves in the insets correspond to the theoretical estimates of Eq. \eqref{eq: amp estimate} (left) and Eq. \eqref{eq: vol estimate} (right), respectively.}\label{fig:resonancebeta}
\end{figure}

In a similar vein, Fig. \ref{fig:resonancew0}(b) shows the concomitant cavity volume in the steady state. Here, we employ the same scaling for the driving frequency, $\Omega/\Omega_0$, but scale the cavity volume $V\Omega_0$. This ensures universal behaviour at low driving frequencies, in good correspondence with the relevant scaling law, shown by the dashed red line. The same universality applies to the theoretical estimate of Eq. \eqref{eq: vol estimate}, which the chosen scaling renders independent of $\Omega_0$, such that the dotted red curve shown in the inset applies to all of the plotted curves.
In contrast, the plotted curves differ by a factor $\Omega_0$ in the high-frequency limit, although they scale identically with $\Omega$. This cannot be scaled out without giving up the universality at low and intermediate frequencies. Consequently, we again find that, first and foremost, increasing $\Omega_0$ shifts the resonance, and suppresses the generation of free volume in the low- and intermediate frequency regimes. This effect vanishes in the high-frequency limit, which is dominated by inertial effects. 

Looking beyond these features, which are absorbed into the chosen scaling, the more subtle effects of varying the network's natural frequency can once again be found around the system's resonance. One trend that is apparent from comparing Figs. \ref{fig:resonancew0}(a) and (b) is that the resonance of the cavity free volume is shifted to higher driving frequencies relative to that of the oscillation amplitude. This stems from the cavity relaxing viscoelastically as a function of time, which biases the results, both in magnitude and in position, toward faster actuation. This effect becomes more pronounced if the natural time scale of the oscillation, $\sqrt{m/k}$, is large relative to the relaxation time of the polymer gel, $\tau$. Hence the effect is most apparent from the silver curve, corresponding to $\Omega_0=\tau\sqrt{k/m}=0.1$.

Following this, the effect of varying the damping ratio, i.e., the friction coefficient of the viscoelastic medium in which the colloid is embedded, follows a very similar trend, as Fig. \ref{fig:resonancebeta} shows. Here, the different curves correspond to $B=0.01$ (black), $B=0.1$ (grey), and $B=1.0$ (silver), and we set $\Omega_0=0.7$. Since the value of $B$ has no effect on the scaling laws and theoretical estimates we use, varying the damping ratio \textit{solely} affects the form of the resonance peak, as highlighted by the insets. 
First and foremost, we see that the resonance in terms of both amplitude (Fig. \ref{fig:resonancebeta}(a)) and cavity volume (Fig. \ref{fig:resonancebeta}(b)) is suppressed upon increasing the damping ratio, as expected. Perhaps more interestingly, we find that for small values of the damping ratio, where the theoretical estimates of Eqs. \eqref{eq: amp estimate} and \eqref{eq: vol estimate} are most suitable, the resonance peak splits in two. We expect that this stems from the interplay between purely elastic interactions and the external field, since increasing $B$ suppresses this split. The same is true for the finer model features apparent in the low-frequency regime, which are, like the split resonance peak, discussed in the Supplementary Material.


\section{Conclusion}\label{sec:conclusion}

In summary, we have investigated the dynamics of a driven colloid embedded in a cross-linked polymer gel, immersed in a viscous background fluid, as a mechanism for rendering such materials responsive. Through a relatively simple, yet non-trivial, model derived from the harmonic oscillator, we show that the oscillatory motion of the colloid can induce a microscopic cavity of a size far exceeding the length scale of the colloid in the steady state. As a function of the driving frequency, we find that the oscillation is well described by simple scaling laws in the low- and high-frequency limits, and we propose closed-form theoretical estimates to describe the resonance that occurs at intermediate frequencies. These estimates qualitatively describe the dependence of the resonance on model parameters, and become quantitative for strongly underdamped systems.

In particular, we conclude that the model dynamics is primarily governed by an interplay between the elastic properties of the network and the external field, with the local friction mainly having a suppressing effect. Specifically, we report that increasing the elastic restoring force acting throughout the network, accessible through the gel's elastic modulus or cross-link density, increases the resonant driving frequency, and diminishes the resonant amplitude and steady-state cavity volume for low and intermediate frequencies; this effect vanishes in the high-frequency limit.
Finally, the model also exhibits a variety of finer features, which we rationalise in the Supplementary Material by considering various simplifying assumptions.

Our work provides a proof of concept for effecting local free-volume generation, and with it responsivity, in cross-linked polymer gels. We note, however, that the same model considerations could be tailored to describe polymer melts. In that case, there is no longer a direct link between the viscosity of the background fluid and the friction coefficient, which then instead derives from the melt itself behaving like a liquid on long time scales. In experiments, such a set-up could be realised by embedding colloidal particles into a cross-linked polymer network or melt during the synthesis. If the colloids are provided with a sufficiently strong coupling to the external field of choice, e.g., electric, we expect oscillatory actuation to give rise to appreciable changes in volume and porosity. 

Although in this paper we focused on the oscillation of a single colloidal particle embedded in a cross-linked polymer gel, we expect our main findings to also be applicable to larger numbers. This is because we expect at least a linear superposition of the individual responses, possibly amplified by their mutual interactions mediated by the polymer network \cite{di2012analytical}. Such an amplification can be explored further by considering shrewdly placing the colloids, e.g., on a lattice, rather than dispersing them randomly in the gel. Regardless of placement, the interactions should be relevant even for relatively small colloid densities, e.g., on length scales far exceeding the cross-link spacing, since such network-mediated interactions decay algebraically in space as $1/r$. Thus, our approach can be interpreted as a lower bound for such systems, in which we effectively describe a small volume element of the material in which a single colloid is embedded, and disregard possible network-mediated interactions. This presents a first step toward rendering arbitrary cross-linked polymer networks suitable for e.g., self-cleaning purposes, pattern formation, and transport of molecular cargo.

The mechanism we highlight here also has broader implications for the liquid crystal networks discussed in the introduction, where reorientation or cis-to-trans isomerisation of rod-like molecules gives rise to the free-volume generation. Our findings suggest that the expansion reported for these materials upon actuation need not be inherently linked to the molecules being grafted to the polymer network \cite{liu2017protruding,gelebart2018photoresponsive}, but can instead follow purely from reorientation inside the polymer matrix. This puts our work in broader perspective.

Finally, a possible extension of our work concerns the assumption of plastic rearrangements occurring in the crosslinked polymer gel. In this work we assumed, for the sake of simplicity, that both elastic relaxation and plastic rearrangements occur on the same time scales. It would be interesting, however, to probe how varying the time scale corresponding to plastic rearrangements, which we expect to be slower than that of elastic relaxation in practice, influences the steady state. Specifically, such an extension may render the model, and in particular the steady state it predicts, less sensitive to the transient stages of dynamics.

\begin{acknowledgments}
	This research received funding from the Dutch Research Council (NWO) in the framework of the ENW PPP Fund for the top sectors and from the Ministry of Economic Affairs in the framework of the `PPS-Toeslagregeling'.
\end{acknowledgments}


\bibliography{main}

\end{document}